\documentclass[10pt,tightenlines,eqsecnum,floats,aps,amsmath,amssymb,showpacs,nofootinbib,superscriptaddress,twocolumn,prd]{revtex4-1}

\newcommand{\comment}[1]{}
\usepackage{array,multirow,graphicx}
\usepackage{enumerate}
\usepackage{graphicx}
\usepackage[table]{xcolor}
\usepackage{euscript,amssymb}
\usepackage{mathrsfs}
\usepackage{setspace}
\usepackage{makecell}
\usepackage{amsfonts}
\usepackage{enumitem}
\usepackage{amsmath}
\usepackage{amsbsy}
\usepackage{mathrsfs}
\usepackage{fancyhdr}
\usepackage{esint}
\usepackage{amsthm}
\usepackage{empheq}
\usepackage{mathtools}
\usepackage{amsbsy}
\usepackage{epsfig}
\usepackage{color}

\usepackage{booktabs}
\usepackage{colordvi}

\DeclareFontFamily{OT1}{pzc}{}
\DeclareFontShape{OT1}{pzc}{m}{it}{<-> s * [1.2] pzcmi7t}{}
\DeclareMathAlphabet{\mathpzc}{OT1}{pzc}{m}{it}
\usepackage[unicode=true, pdfusetitle,
bookmarks=true,bookmarksnumbered=false,bookmarksopen=false,
breaklinks=false,pdfborder={0 0 1},backref=false,colorlinks=false]
{hyperref}
\usepackage{graphicx}
\usepackage{xcolor,fontawesome}

\usepackage{amsfonts}
\usepackage{enumitem}
\usepackage{amsmath}

\usepackage{amssymb}

\usepackage{colordvi}
\usepackage{fancyhdr}
\usepackage{esint}

\begin{document}

\title{Exploring the cosmic censorship conjecture with a Gauss-Bonnet sector}

\author{Yaser Tavakoli}
\email{yaser.tavakoli@guilan.ac.ir} 
\affiliation{Department of Physics,
University of Guilan, Namjoo Blv.,
41335-1914 Rasht, Iran}
\affiliation{School of Astronomy, Institute for Research in Fundamental Sciences (IPM),	P.O. Box 19395-5531, Tehran, Iran}

\author{Ahad K. Ardabili}
\email{ahad.ardabili@altinbas.edu.tr} \affiliation{Department of Basic Sciences, Alt{\i}nba\c{s} University, 34217 Istanbul,  Turkey}

\author{Paulo Vargas Moniz}
\email{pmoniz@ubi.pt} \affiliation{CMA-UBI and Departamento de F\'{\i}sica, Universidade da Beira Interior, 6200 Covilh\~{a}, Portugal}

\date{\today}


\begin{abstract}
	The Dvali-Gabadadze-Porrati (DGP) braneworld model is 
	employed to study the gravitational collapse of dust, with a Gauss-Bonnet (GB) term present in the five-dimensional bulk. We
	find that, within the normal (nonself-accelerating) DGP branch and due to the curvature effects from the  GB component on the brane,
	the black hole  singularity  acquires 	modified features.
	More precisely, during collapse and for	a finite comoving time, 
	before a  singularity would emerge at the zero physical radius,	the first time derivative of the Hubble rate  diverges,  whereas the brane energy density and the Hubble rate remain finite. This is a peculiar behavior	which displays similar properties
	to the  sudden singularity occurring in 
	particular late-time cosmological frameworks. Furthermore, the question of whether this 
	altered singularity can be  viewed by an 
	external observer or will be hidden by a black hole horizon
	is addressed. We establish
	that,  depending on the given induced-gravity parameter and the GB coupling constant, there exists a {\em threshold mass} for the collapsing dust, below which no trapped surfaces evolve as the collapse proceeds toward the singularity. In other words, a {\em naked  sudden singularity} may form.
\end{abstract}


\maketitle

\section{Introduction}

Recent astrophysical evidence including the first image of 
a black hole \cite{akiyama:2019first} and gravitational wave detection \cite{Abbott:2016blz,abbott:2016properties,Abbott:2016nmj} 
constitute remarkable achievements. In particular, they have the potential to enhance immensely the study of 
phenomena with large masses producing events bearing startling 
energy ranges. A
strong  gravitational field, governed by the  theory of general relativity (GR), plays a dominant role,
in so far as 
thorough analysis has demonstrated. But can such observational procedures 
advise us into selecting paths to  further probe the nature of our universe?
Concretely,  identifying on the one hand, domains where to defy our appraisal of gravitational collapse. And, on the other hand, suggesting therefore whether there is a broader 
gravitational theory beyond GR.

The process of the gravitational collapse of a sufficiently  massive star leads to the formation of a spacetime singularity \cite{Hawking:1973uf,Hawking:1969sw}. These are regions with zero physical radii where the energy density and the curvature of spacetime blow up. 
  Under such extreme physical conditions,
extending them toward and in the Planck regime,
we expect that the theory of GR 
must be  replaced
by a more complete theory of gravity.
This could require to find a quantum theory of gravity which modifies the spacetime geometry 
 at  Planck  ranges \cite{Bojowald:2005qw,Bojowald:2001xe,Bojowald:2007ky,Ashtekar:2008ay,Horowitz:1989bv}. 
In past decades there has been huge effort toward formulating a theory to be valid in very small length scales and high energy density. 
Accordingly, it would be of interest to examine the 
presence or otherwise, the
absence of spacetime singularities in these theories. Moreover, it would be pertinent to investigate if and how a new theory of 
gravity,  incorporating  geometrical sectors other than GR terms, would modify the nature of any such singularities.

An idealized model of gravitational collapse in GR was studied by Oppenheimer and Snyder in 1939  which corresponds to the collapse of a homogeneous spherical dust cloud \cite{Oppenheimer:1939ue}.
In this model, the end product of the collapse process is a black hole, that is, after a concrete time  elapses
from the initial configuration, the collapsing star proceeds toward the horizon and it continues until a spacetime singularity is formed. Therefore, the final  singularity will be hidden behind the Schwarzschild horizon. 
This model for gravitational collapse, although treating only a very simplified situation, was highly important for providing the intuition  to the more general problems such as the concept of the trapped surface which plays a central role in the Hawking and Penrose singularity theorems  \cite{Hawking:1973uf}. 
This is indeed the weaker form of the so-called cosmic censorship  conjecture (CCC).  Nevertheless,  a stronger form of the CCC was introduced by Penrose   which states that any singularities that arise from regular initial data are not even locally visible \cite{penrose:1965zero,Penrose:1969pc,Penrose:1964wq}. In other words, it is not possible for future-directed null geodesics, originating from vicinity of the singularities, to escape to infinity leaving the center  visible to outside observers (i.e. by forming a {\em naked singularity}). This stronger conjecture is verified in the Oppenheimer-Snyder model as mentioned above.
For particular matter distributions, and for various models of gravity, the fate of CCC and its possible violation  has been extensively studied  in the literature \cite{Choptuik:1992jv,Christodoulou:1994hg, Lemos:1991uz,Brassel:2019bam,Zeng:2019hux,Ziaie:2019klz,Tavakoli:2015llr} (see also \cite{joshi_2007, Joshi:1993zg, Goswami:2003bs,Lemos:1997bd,Goswami:2004ne}).
Hence, exploring the scope of that conjecture within a new theory of gravity is a valuable purpose.

A significant scope of approaches, within either astrophysical and cosmological settings, have contributed to the literature of modified gravity and singularities, for example  \cite{Easson:2003ia,Das:2006pw,Ashtekar:2006uz, Ashtekar:2006rx, Tavakoli:2013rna, Tavakoli:2013tpa, BouhmadiLopez:2009jk,Bouhmadi-Lopez:2017kvc,Tavakoli:2015llr,Bouhmadi-Lopez:2014cca,Marto:2013soa}.
Noteworthy,
string and $M$ theories are two  of the main candidates  to unify quantum mechanics and GR \cite{greensuperstring,Townsend:1996xj,Duff:1996aw}. 
It is of relevance to point that
the low energy limit of these higher dimensional theories exist in a four-dimensional (4D) world.  Inspired by these theories, braneworld models were proposed in which the observable world is a 4D brane embedded in a 5D bulk \cite{Deffayet:2000uy, Dvali:2000hr, Cho:2001nf, Randall:1999ee, Randall:1999vf}.  
These models are characterized by the feature that standard 
matter fields are confined to the $(3+1)$D brane whereas gravity propagates in the higher dimensional bulk. Moreover, gravity appears 4D on the brane via the warping of the bulk dimension. In other words, from the point of view of the brane observer, gravity appears 5D at high energy whereas it is revealed 4D in the low energy regime.

An interesting  braneworld scenario was subsequently
suggested by Dvali, Gabadadze
and Porrati (DGP) \cite{Dvali:2000hr}, which assumes an Einstein-Hilbert brane action where the brane Ricci scalar can be interpreted as arising from a quantum effect due to the interaction between the bulk gravitons and the matter on the brane \cite{Dvali:2000hr, Deffayet:2000uy}. 
In this model, the induced gravity term on the brane dominates
below a certain ``cross-over'' scale $r_c$, so gravity becomes 4D.
However, at scales greater than $r_c$, gravity ``leaks off'' the brane (the 5D Ricci term in the DGP action begins to dominate over the 4D Ricci term) so it appears to be 5D to observers on the brane. Thereby, the bulk is no longer required to be warped in the DGP  model, so that the DGP brane is instead embedded  in an infinite dimensional Minkowski bulk. 
Therefore, the DGP braneworld model modifies the low energy regime of a gravitational theory (i.e., IR modifications). From a geometrical point of view, adding 
a (UV-like) Gauss-Bonnet (GB) term in 5D  is natural:  
the most general unique action in 5D must include a GB sector \cite{Kofinas:2003rz, Brown:2005ug, BouhmadiLopez:2009jk, BouhmadiLopez:2008nf}. This geometrical term thus modifies gravity in the high energy regime.

Within the context of the above paragraphs, our objective herein this paper is to apply a specific IR-UV combination to gravity to 
investigate the fate of gravitational collapse for a specific modified 
gravity  setting  and to understand the consequences as far as CCC is concerned
(see for example \cite{Creek:2006je,Chakraborty:2015taq}). We thus consider a DGP-GB 
brane model where a generic barotropic fluid  is present in  the brane,  with  the bulk
characterized by
higher curvature effects from GB terms.
We will show that 
such setting contains interesting 
{\it new} type of (gravitational collapse) singularities  (see e.g., \cite{Barrow:2004xh,FernandezJambrina:2004yy, BouhmadiLopez:2009jk}).
In addition,   the induced-gravity parameter and the GB coupling constant, comply to bring about a {\em threshold mass}. Depending on its value,  trapped surfaces may evolve as the collapse proceeds or, instead  a naked singularity may form. This can be of relevance within the scope of appraising the CCC within a given IR-UV combination as means to explore modified  gravity.
We are aware that ours is a particularly simple  model  within a modified gravity theory. That notwithstanding,  we are confident that by  means of the ingredients we add, together with the subsequent results we report herewith, this will assist our arguments and further understanding of the CCC,  plus of the configurations that either support it or suggesting its violation.

We thus organize this paper as follows. In Sec.~\ref{model}, 
we describe a region of homogeneous distribution of  a barotropic fluid undergoing isotropic and homogeneous collapse on the DGP brane (with a GB term). Then,  in Sec.~\ref{model-solutions} we will present different classes of solutions to the modified Friedmann equation in the internal collapsing region, where the physical reasonability  of the solutions will be thoroughly analyzed. In Sec.~\ref{Sec:singularity}, by selecting the 
relevant solutions for gravitational collapse,  we will investigate the fate of the  singularity  emerging on the brane. We will show that a GR  singularity will be 
replaced by a different type, 
which we label as {\it sudden} (naked/black hole)  singularity.  
In Sec.~\ref{Sec:Exterior-geometry}, by applying appropriate junction conditions on the fluid boundary, we will  analyze the evolution of trapped surfaces and will discuss the possible candidates for the exterior spacetime geometry both for the formation of a black hole and of a naked singularity Finally, in Sec.~\ref{conclusion} we will present the conclusion and discussions concerning our work.

\section{Gravitational collapse in the DGP-GB model}
\label{model}

Different classes of collapsing fluid models have been studied in the context of GR, which result in a naked singularity as collapse end state \cite{Goswami:2005fu, Goswami:2004ne, Giambo:2005se}. It turns out that  the collapse  outcomes  depend on the boundary conditions and the initial fluid profile. 
Our purpose in this section is to construct a class of continual collapse model in a braneworld scenario with an interior region of generic barotropic fluid, such that the interplay between the brane and  bulk (which is modified due to the higher order curvature terms) could alter the nature of singularity or/and development of trapped surfaces in the spacetime. We  require that the weak and null energy conditions are preserved throughout the collapse, though the emergent effective pressure may be negative in the vicinity of the singularity. 
To this aim, we introduce a {\em region} of  gravitational collapse of a  perfect fluid localized on a DGP brane while a GB term is considered on the bulk. We first extract the dynamical equations for the  collapse scenario (in the interior spacetime), and then, we will discuss the physical reasonability of the model by analysing the energy conditions.

\subsection{The model}

The general gravitational action  with induced gravity on the ($\mathbb{Z}_{2}$ symmetric) brane (DGP term)  and quadratic  GB term is 
\begin{align}
S &= \frac{1}{2} \int dx^5 \sqrt{- g^{(5)}} \, M^3_{(5)} \left[{\cal R}_{(5)}  +\alpha\mathcal{L}_{\rm GB}\right] \nonumber \\
& \quad \quad  + \frac{1}{2} \int dx^4 \sqrt{- g} \, M^2_{(4)} \left({\cal R}_{(4)} + {\cal L}_{\rm matt} \right),
\label{action}
\end{align}
where $\mathcal{L}_{\rm GB}$ is the Lagrangian density of the GB sector:
\begin{align}
\mathcal{L}_{\rm GB} =  {\cal R}_{(5)}^2  - 4{\cal R}_{(5)ab}{\cal R}_{(5)}^{ab}    +  {\cal R}_{(5)abcd}{\cal R}_{(5)}^{abcd}\,   ,
\end{align}
and $\alpha(\geq 0)$ is  the coupling constant  associated to the GB term  in the   Minkowski bulk. 
The GB term corresponds to  the leading order quantum corrections to gravity in an effective action approach of string theory \cite{Gross:1986mw}, where $\alpha=1/8g_s^2$ is related to string scale $g_s^2$ and can be identified with the inverse of string tension \cite{Charmousis:2002rc}.
The case $\alpha=0$ corresponds to the standard DGP model \cite{Dvali:2000hr, Deffayet:2000uy}. Moreover, $M_{(4)}$ and $M_{(5)}$ are the fundamental  four and five dimensional Planck masses, respectively.

By considering a homogeneous, isotropic barotropic fluid for the collapsing region on the brane, the interior spacetime can have a FLRW metric. In particular, we will consider an internal spacetime model on the brane given by a marginally bound ($k = 0$) case, i.e.  the initial configuration of the fluid is assumed to be located at very large physical radius. Thus, the interior metric is of the form \cite{Lemos:1997bd,Goswami:2003bs}
\begin{equation}
ds^2=-dt^2+a^2(t)\left(dr^2+r^2d\Omega^2\right),
\label{metric-interior}
\end{equation}
where $d\Omega^2$ denotes the line element on a two-sphere.
The  generalized Friedmann equation for the brane, equipped with the metric (\ref{metric-interior}),  then reads \cite{Kofinas:2003rz}
\begin{equation}
H^2\, =\, \frac{\kappa_{(5)}^2}{6r_c}\rho \pm
\frac{1}{r_c}\left(1+\frac83\alpha H^2\right)H,
\label{modifiedfriedmann}
\end{equation}
where $\kappa_{(4)}^2 = M_{(4)}^{-2} $  and  $\kappa_{(5)}^2 = M_{(5)}^{-3}$ are four  and  five-dimensional gravitational constants, respectively; and $r_c$ is the induced-gravity ``cross-over'' length scale  which marks the transition from 4D to 5D gravity:
\begin{equation}
 r_c=\kappa^2_{(5)}/2\kappa^2_{(4)}.
\end{equation}  
The two signs on the right-hand side of Eq.~(\ref{modifiedfriedmann}) are associated with  two branches DGP$(\pm)$, in the case $\alpha=0$, which  correspond to different embeddings of the brane in the Minkowski bulk.
Moreover, $H$ is the Hubble parameter  of the collapsing region on the brane, whose matter content is a {\em barotropic fluid} with $\rho$ representing its total energy density.

We intend to inquire the effects of the  UV modifications to the collapse dynamics on the brane, with a particular interest on the influence of the GB term  on the bulk. To be more generic, let us for the moment,   consider a perfect fluid with an energy density $\rho$, a pressure $p$ and an equation of state parameter $w>-1$ (where $p=w\rho$), for the collapse matter field. In the DGP model, the energy-momentum tensor for the localized matter field on the brane is conserved as in GR  \cite{Deffayet:2000uy,Deffayet:2002fn},  which implies that $\dot{\rho}+3H(\rho+p)=0$. Thus, the   energy density of the collapse matter  can be written as 
\begin{equation}
    \rho=\rho_{0}\left(\frac{a_0}{a}\right)^{3(1+w)},
    \label{energy-dust}
\end{equation}
where, $a_0$ and $\rho_0$ are respectively, the initial values of the scale factor and the energy density of the collapsing region on the brane\footnote{The explicit form of  the matter energy density (\ref{energy-dust}), i.e. the particular value of $w$ here, follows the dynamics of the collapse on the brane due to the Friedmann equation (\ref{modifiedfriedmann}). Nevertheless, in the next section, we will find the Hubble rate solutions to Eq.~(\ref{modifiedfriedmann}) as generic functions of  $\rho$. Therefore, for a desirable value of $w$, it will be straightforward to evaluate the associated Hubble rate capturing the collapse dynamics on the brane.}.
On the other hand, the modified Raychaudhuri equation is obtained easily from the time derivation of the Friedmann equation (\ref{modifiedfriedmann}) on the brane and the conservation of the brane energy density as
\begin{equation}
\dot{H}  = - \frac{\kappa_{(4)}^2 r_c H(1+w)\rho}{2r_cH-\epsilon(1+8\alpha H^2)}\, ,
\label{Eq:Raychaudhuri-1}
\end{equation}
in which a dot stands for a derivative with respect to the proper time on the brane.
Moreover, $\epsilon=+1, -1$ in Eq.~(\ref{modifiedfriedmann}) correspond, respectively, to the ($+$) sign, which is usually called (in the cosmological context) as the {\em self-accelerating branch}, and the ($-$) sign, which is called the {\em normal branch}.

To deal with the Friedman equation  it is helpful to  introduce  the  following dimensionless variables:
 \begin{eqnarray}
\bar H  := \frac83 \frac{\alpha}{r_c}H , \quad  \bar\rho  := \frac{32}{27}
\frac{\kappa_{(5)}^2\alpha^2}{r_c^3} \rho , \quad  b := \frac83\frac{\alpha}{r_c^2} \, . \quad  \label{eq8a}
\end{eqnarray}
Then, we can express the  Friedmann  equation  (\ref{modifiedfriedmann}) as
\begin{eqnarray}
{\bar H}^2={\bar \rho} + \epsilon(b{\bar H}+{\bar H}^3), \label{Friedmannn-epsilon} 
\end{eqnarray} 
Note that, the two cubic equations can be transformed to each other by changing the sign of $\bar H$, therefore, the solution to the $(+)$ branch is simply equal to the negative of the solution to the  $(-)$ branch.

In the standard DGP context (as in the GR), as the collapse proceeds on the brane, its energy density and Hubble rate (i.e., solutions to Eq.~(\ref{modifiedfriedmann}) for $\alpha=0$) grow and finally diverge as the scale factor vanishes [cf. Eq.~(\ref{DGP-pure-sol})]. Nevertheless, as we will show in Sec.~\ref{Sec:singularity}, for $\alpha\neq 0$ a modified dynamics for the collapsing region emerges at the brane whose behavior is different from that of the real matter field (on the brane equipped with the energy density (\ref{energy-dust})) in the standard DGP or GR models.  This  behavior is a consequence of the extra-dimension which is based in mapping the brane evolution to that of an equivalent 4D general relativistic collapse model with a modified matter content \cite{Sahni:2002dx,Lue:2004za,BouhmadiLopez:2009jk,Bruni:2001fd}. 
	More precisely, a 4D Einstein tensor on the brane is derived from an induced metric (using the junction conditions on the brane hypersurface embedded in the Minkowski bulk) through the brane Einstein-Hilbert action \cite{Dvali:2000hr, Deffayet:2000uy, Deffayet:2001pu, Deffayet:2002fn, Kofinas:2003rz}. 
	Therefore a 4D gravitational field equations emerge on the brane 
	with an effective energy-momentum tensor which  corresponds to a composition of the real matter energy-momentum tensor (localized on the brane) and the geometrical sector associated to the junction conditions on the 4D brane embedded in the 5D bulk, being encoded on the Hubble rate [cf. Eq.~(\ref{energydensity-1})] \cite{Kofinas:2003rz}. 
	The geometrical sector contains the effects of both the induced metric and the Lovelock terms due to GB action in the bulk. It was shown that, using the Gauss-Codazzi equations, the conservation equation for the real matter energy-momentum tensor and the Bianchi identities on the brane are both satisfied \cite{Kofinas:2003rz, Deffayet:2002fn, Germani:2002pt, Davis:2002gn, Maeda:2003ar,Shiromizu:1999wj}.

Therefore, in order to get the desired evolution of the effective energy density which alters the collapse fate, it is necessary to solve the cubic Friedmann equation (\ref{Friedmannn-epsilon}) on a suitable branch (i.e., the $(-)$ branch as we will discuss in the sequel). We will present this procedure on the next section.

\subsection{Energy conditions}  

For a desired branch, once the mathematical solutions to the modified Friedmann equation (\ref{Friedmannn-epsilon}) are known, one  needs to discuss their physical implications in the collapse scenario. In particular, to assure physical  reasonability of the solutions,  a set of energy conditions must be satisfied by the collapsing system \cite{Hawking:1973uf}. 
There are two important energy conditions to be satisfied by all (classical) matter fields. 
The first  is the \textit{weak energy condition} (WEC),  which requires that  the energy density  measured by any local timelike observer must be  non-negative. And the second one is the \textit{null energy condition} (NEC), which ensures that the flow of the energy for any timelike observer is not spacelike (i.e. the speed of the energy flow does not exceed the speed of the light).

In order to study the energy conditions in their well-known general relativistic form, it is convenient to introduce an effective energy density $\rho_{\rm eff}$ on the brane  which obeys the equation of state of a perfect fluid, equipped with an effective parameter $w_{\rm eff}$ and an effective pressure $p_{\rm eff}\equiv w_{\rm eff}\rho_{\rm eff}$. For a given $\epsilon$, we thus rewrite  the modified Friedmann equation (\ref{modifiedfriedmann}) based on these effective quantities as
\begin{eqnarray}
H^2 &=& \frac{\kappa_{(4)}^2}{3}\rho_{\rm eff} ,
\label{modifiedfriedmann-2}
\end{eqnarray}
where  $\rho_{\rm eff}$ is defined as
\begin{eqnarray}
\rho_{\rm eff} &:=& \rho +
\frac{6}{\kappa_{(5)}^2}\epsilon \left(1+\frac83\alpha H^2\right)H .
\label{energydensity-1}
\end{eqnarray}
This effective energy density and its corresponding pressure should satisfy an effective conservation equation  as follow:
\begin{equation}
\dot{\rho}_{\rm eff}+3H \left(\rho_{\rm eff}+p_{\rm eff}\right) =0,
\label{eff-conservation}
\end{equation}
Then, $\dot{\rho}_{\rm eff}$ is obtained by taking a time derivative of  Eq.~(\ref{energydensity-1}), as
\begin{eqnarray}
\dot{\rho}_{\rm eff} = \dot{\rho} + 
\frac{6}{\kappa_{(5)}^2}\epsilon\left(1+8\alpha H^2\right)\dot{H}.
\label{energydensity-der}
\end{eqnarray}
By setting Eq.~(\ref{energydensity-der}) into the conservation equation (\ref{eff-conservation}), 
we  find the effective pressure  $p_{\rm eff}$  as
\begin{eqnarray}
p_{\rm eff} = -\rho_{\rm eff} + \frac{2r_cH\rho}{2r_cH - \epsilon(1+8\alpha	H^2)}  \, .
\label{pressure-eff}
\end{eqnarray}
Note that, Eq.~(\ref{pressure-eff}) is equivalent to an effective Raychaudhuri equation consistent  with Eq.~(\ref{Eq:Raychaudhuri-1}):
\begin{equation}
    \dot{H} = -\frac{\kappa_{(4)}^2}{2}(\rho_{\rm eff}+p_{\rm eff}).
\end{equation}
It is then clear that, when the effective matter on the brane is modified, due to curvature term on the bulk, the effective equation of state on the brane (in the GR picture) becomes different from the real (collapse) matter content of the brane. More precisely, the effective equation of state now becomes
\begin{eqnarray}
w_{\rm eff} = -1 + \frac{2r_cH(1+w)}{2r_cH - \epsilon(1+8\alpha	H^2)}\frac{\rho}{\rho_{\rm eff}}  \, .
\label{EoS-eff}
\end{eqnarray}

Now that we have the effective energy density and the effective pressure, we can write the WEC and the NEC for the effective energy-momentum tensor as:
\begin{align}
\text{WEC}:\quad  &\rho_{\rm eff} =
\rho +
 \frac{6}{\kappa_{(5)}^2}\epsilon\left(1+\frac83\alpha H^2\right)H\geq 0, \\
\text{NEC}: \quad  &\rho_{\rm eff}+p_{\rm eff}   =
\frac{2r_cH\rho}{2r_cH -\epsilon(1+8\alpha
	H^2)} \geq 0 \, .
\end{align}
To simplify our analysis of the energy conditions, we rewrite the above equations in terms of the dimensionless parameters:
\begin{align}
&
\textrm{WEC}: \quad \quad \bar\rho + 
\epsilon \left(b+  \bar{H}^2\right)\bar{H}\geq 0, \label{EC1} \\
& 
\textrm{NEC}: \quad \quad \frac{2\bar{H} \bar\rho}{2 \bar{H} - 
	\epsilon(b+ 3\bar{H}^2)} \geq 0  .\label{EC2} 
\end{align}
The WEC can be examined by comparing Eq.~(\ref{EC1}) with the Friedmann equation  (\ref{Friedmannn-epsilon}), noting that 
\begin{equation}
 \bar\rho + 
\epsilon \left(b+  \bar{H}^2\right)\bar{H} = \bar{H}^2\geq 0.
\end{equation}
Therefore, the WEC  always holds. 
For the  NEC to be satisfied, the denominator of the Eq.~(\ref{EC2}) should be always negative. This is equivalent to the condition below:
\begin{equation}
2 \bar{H} - \epsilon 
(b+ 3\bar{H}^2)<0. \label{NEC}
\end{equation}
In the above discussion, we have followed the fact that $\rho>0$ for the energy density of the real matter on the brane,  whereas  $\bar{H}<0$  guarantees a collapsing process.

For the $(-)$ branch, Eq.~(\ref{NEC}) is satisfied  if,
 \begin{equation}
 	  \bar{\mathscr{H}}^{(-)}_1<\bar{H}<\bar{\mathscr{H}}_{2}^{(-)},
 	  \label{NEC-negative}
\end{equation}
where
\begin{align}
 	\bar{\mathscr{H}}^{(-)}_1\, &=\,  -\frac{1}{3}(1+\sqrt{1-3b}), \label{EC2b0}  \\ 
 	\bar{\mathscr{H}}^{(-)}_2\, &=\,  -\frac{1}{3}(1-\sqrt{1-3b}).
 	\label{EC2b}
\end{align}
Similarly, for the $(+)$ branch, the condition (\ref{NEC}) implies that
\begin{equation} 
\bar{H}<\bar{\mathscr{H}}^{(+)}_2
\quad \quad \text{or} \quad \quad \bar{H}>\bar{\mathscr{H}}^{(+)}_1,
\label{NEC-positive}
\end{equation}
where
\begin{align}
 	\bar{\mathscr{H}}^{(+)}_1\, &=\,  \frac{1}{3}(1+\sqrt{1-3b}),  \\ 
 	\bar{\mathscr{H}}^{(+)}_2\, &=\,  \frac{1}{3}(1-\sqrt{1-3b}).
	\label{EC2c}
\end{align}
The first condition is satisfied because for a collapse process we always have  $\bar{H}<0$. However, the second condition is not the case in the present context.

From the NEC (\ref{NEC-negative})  for the ($-$) branch, it is clear that the Hubble rate  is limited by the two values  $\bar{\mathscr{H}}^{(-)}_1,~\bar{\mathscr{H}}^{(-)}_2<0$. In contrary, in the $(+)$ branch, the NEC (\ref{NEC-positive}) does not provide a lower limit for $\bar{H}$. This  permits the  divergence of $\bar{H}$ (with negative sign) at the late time stages of the collapse, which subsequently violates the WEC (\ref{EC1}). 
We  are interested in a situation where,  without considering any matter that violates the NEC on the brane, the effective energy density (\ref{energydensity-1}) will behave like that of a matter component with $w_{\rm eff}\leq-1$ on the brane  (that depends explicitly on $\bar{H}$). This would smoothen the divergent nature of the real matter fluid with $w>-1$
at the collapse end state (cf. \cite{Oppenheimer:1939ue,Christodoulou:1984djm, joshi_2007}). We, henceforth in the rest of this paper, will only  consider a $(-)$ branch.

In the next section we will analyse the solutions to the Friedmann equation (\ref{Friedmannn-epsilon}) for the $(-)$ branch and study their physical implications for different collapse scenarios.

\section{The solutions}
\label{model-solutions}

For the $(-)$ branch, the (dimensionless) Friedmann equation (\ref{Friedmannn-epsilon}) can be written  as
\begin{eqnarray}
{\bar H}^3+{\bar H}^2+b\bar H-\bar\rho=0. \label{Friedmannnb} 
\end{eqnarray}
 The number of the solutions to the cubic equations (\ref{Friedmannnb}) depends on the sign of the discriminant function $N$, defined as
\begin{equation}\label{eq10}
N=Q^3+S^2,
\end{equation}
where $Q$ and $S$ are,
\begin{equation}
Q:=\frac{1}{3}\left(b-\frac{1}{3}\right),
\quad \quad  S:=\frac{1}{6}b+\frac{1}{2}\bar{\rho}-\frac{1}{27} \, .
\label{eq11}
\end{equation}
If $N>0$, then we have  three roots, one of which is  real  and the other two are complex conjugates.  For $N = 0$, all of the solutions are real and two of them are equal. And finally if $N<0$,  all roots are real \cite{Abramowitz:1965}.

To simplify the  analysis of the sign of $N$  we rewrite it in terms of two new variables $\rho_{+}$ and $\rho_{-}$ as \cite{BouhmadiLopez:2009jk}
\begin{equation}
N=\frac{1}{4}(\bar{\rho}-\bar{\rho}_{+})(\bar{\rho}-\bar{\rho}_{-}),
\label{eq12}
\end{equation}
where 
\begin{eqnarray}
\bar{\rho}_{+}\, &:=&\, \frac{2}{27}\left[1+\sqrt{(1-3b)^3}\right]-\frac{b}{3}\, ,
\label{rhoone}
\\
\bar{\rho}_{-}\, &:=&\, \frac{2}{27}\left[1-\sqrt{(1-3b)^3}\right]-\frac{b}{3}\, .
\label{rhotwo}
\end{eqnarray}
Now we look at the  behavior of $\bar{\rho}_{\pm}$ with respect to the parameter $b$. If $b$ takes a value  in the  range   $0<b<1/4$, $\bar{\rho}_{+}$ is always positive while $\bar{\rho}_{-}$ is always negative.  Then depending on the relative values of $\bar{\rho}_{+}$ and  $\bar{\rho}$ there are three cases: 
\begin{enumerate}[label=\roman*)]
\item for
$\bar{\rho}>\bar{\rho}_{+} \,\, \Rightarrow \,\, N>0$: \quad ``high energy regime'';
\item
for $\bar{\rho}=\bar{\rho}_{+}  \,\,  \Rightarrow \,\, N=0$: \quad ``limiting regime'';
\item
for $\bar{\rho}<\bar{\rho}_{+} \,\,  \Rightarrow \,\,  N<0$: \quad ``low energy regime''. 
\end{enumerate}
For the range of parameter $b \geq 1/4$,~  $N$ is always positive and there exists a unique real solution.

\subsection{The case $0<b<\frac{1}{4}$}
\label{A}

As mentioned earlier, here we will have three different cases depending on the relative values of the energy density $\bar{\rho}$ of the brane. However, the energy density is time-dependent (it blueshifts in time; i.e. the energy density grows  as the collapse proceeds) so that depending on the initial energy density $\bar{\rho}_{0}$ of the brane, defined by  
\begin{eqnarray}
\bar\rho_0 \, :=\,  \frac{32}{27}
\frac{\kappa_{(5)}^2\alpha^2}{r_c^3} \rho_0 = \frac{\kappa_{(4)}^2}{3} b^2r_c^2
\rho_0 \, ,
\label{energy-trans}
\end{eqnarray}
$N$ and consequently  the solutions to the Hubble rate  will change.

\subsubsection{The initial condition $\bar\rho_0>\bar{\rho}_{+}$}

This corresponds to a case where the brane starts its evolution initially in the high energy regime. Since $\bar\rho$ is an incremental function of time,  $\bar{\rho}>\bar{\rho}_0>\bar{\rho}_+$, thus the brane remains in this regime in the rest of its  progress.
In this case  $N>0$ and we will have a unique (real)  solution:
\begin{equation}
\bar{H}_1(\eta) = \frac{1}{3}\left[2\sqrt{1-3b}\cosh\left(\frac{\eta}{3}\right)-1\right],
\label{EXP1}
\end{equation}
where we have defined the parameter $\eta$ as
\begin{equation}
\quad  \cosh(\eta) \coloneqq \frac{S}{\sqrt{-Q^3}}~, \quad \quad \sinh(\eta) \coloneqq
\sqrt{\frac{N}{-Q^3}} \, .
\label{eq16}
\end{equation}
Note that this solution is always positive so it does not represent a collapsing scenario. Therefore,  this initial condition, i.e. $\bar\rho_0>\bar{\rho}_{+}$,  is not  physically relevant for our study.

\subsubsection{The initial condition  $\bar\rho_0 \leq \bar{\rho}_{+}$}

\begin{figure*}
	\begin{center}
	\includegraphics[scale=0.7]{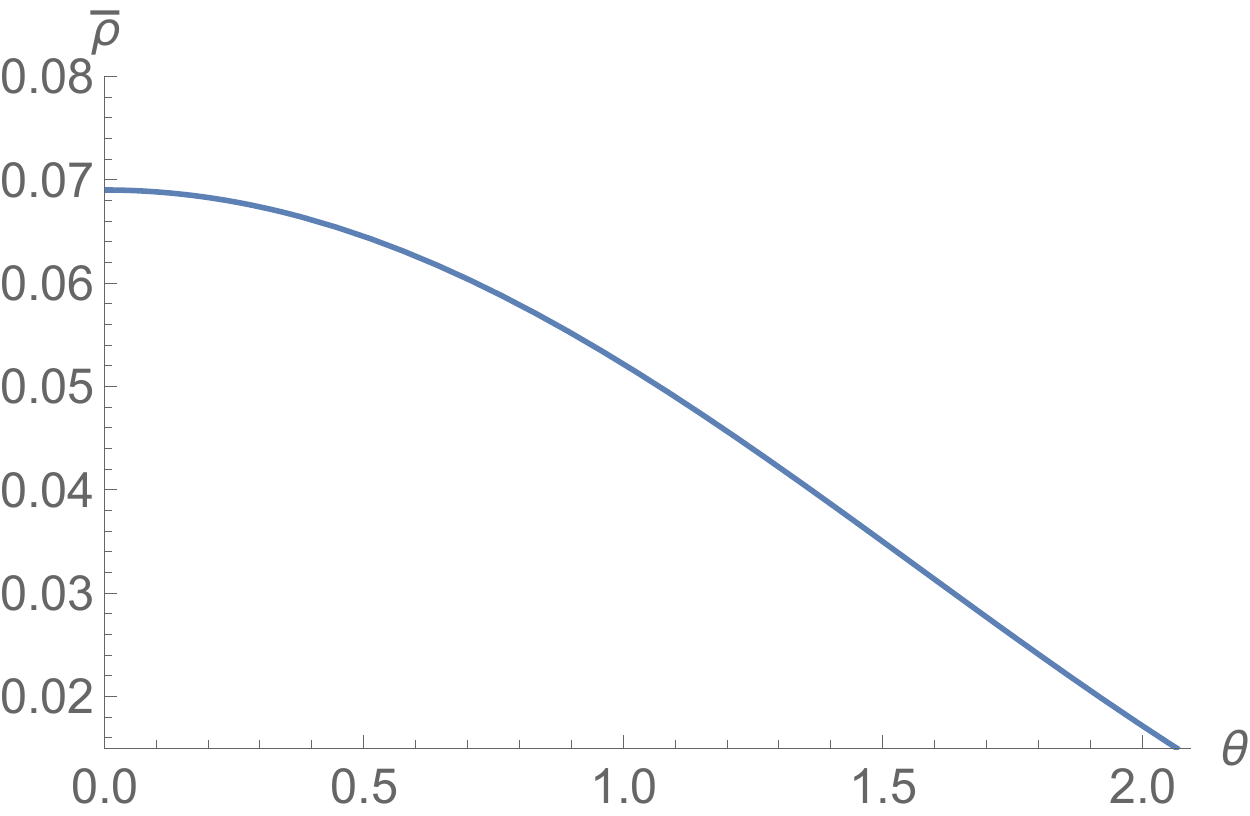}\quad \quad \quad \quad 
		\includegraphics[scale=0.6]{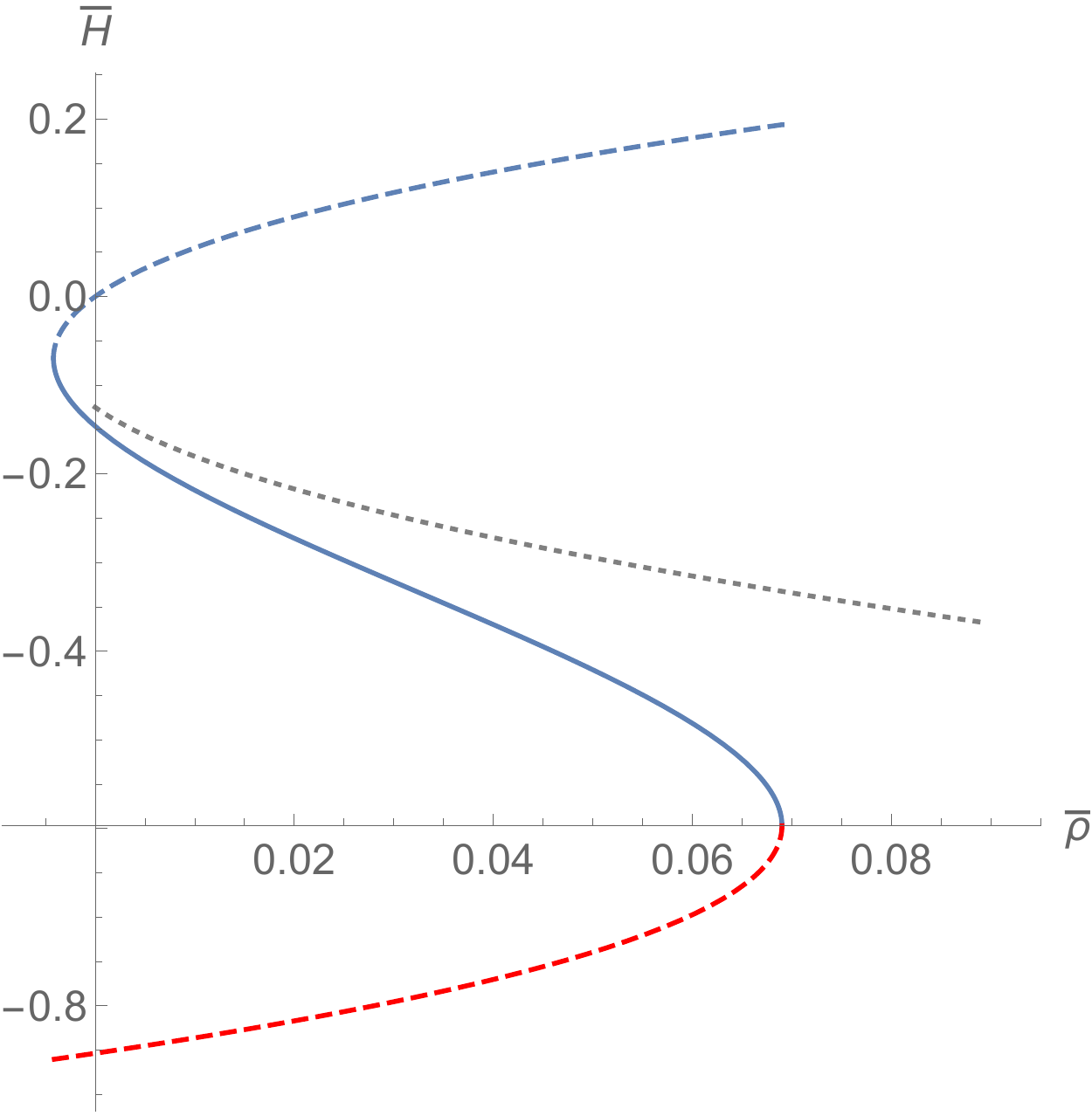}
		\caption{Left plot: the evolution of the energy density (\ref{EnergyDensity1}) of the brane in the range $0\leq \theta\leq \theta_0$ with $\theta_0=2\pi/3$. Right plot: the evolution of the (dimensionless) Hubble rate solutions  $\bar{H}_1(\bar{\rho})$ (solid blue curve), $\bar{H}_2(\bar{\rho})$ (dashed red curve) and $\bar{H}_3(\bar{\rho})$ (dashed blue curve), in the low energy  regime (where $0.014\leq\bar{\rho}\leq 0.069$), within ``DGP-GB'' model. Moreover, the gray dotted line represents the divergent behavior of the Hubble parameter (\ref{DGP-pure-sol}), in a ``standard DGP'' model, in terms of dimensionless parameters.}  \label{Fig:LowEnergyRegime}
	\end{center}
\end{figure*}

This case represents a collapse scenario initiated in a low energy regime. Then, as the brane energy density blueshifts, the collapse can evolve through all three regimes. In the following we will analyse these stages:
\begin{enumerate}[label=(\roman*)]
\item\label{lowEnergy-item} As long as the energy density of the brane is in the range  $\bar{\rho}_0<\bar{\rho}<\bar{\rho}_{+}$, i.e. the brane evolves  in the  low energy regime, there exist three different solutions for the Hubble rate:
\begin{align}
\quad \quad  \bar{H}_{1}(\theta) &=-\frac{1}{3}\left[2\sqrt{1-3b}\cos\left(\frac{\theta+\pi}{3}\right)+1\right],\label{lowenergy1}
\\
\quad \quad  \bar{H}_{2}(\theta) &=\frac{1}{3}\left[2\sqrt{1-3b}\cos\left(\frac{\theta}{3}\right)-1\right],\label{lowenergy2}\\
\quad \quad  \bar{H}_{3}(\theta) &=-\frac{1}{3}\left[2\sqrt{1-3b}\cos\left(\frac{\theta-\pi}{3}\right)+1\right],
\label{lowenergy3}
\end{align}
where, $0<\theta \leq\theta_0$ is defined  as follows:
\begin{equation}
\quad \quad  \cos(\theta) \coloneqq \frac{S}{\sqrt{-Q^3}}\, , \quad \quad \sin(\theta) \coloneqq \sqrt{\frac{N}{Q^3}}\, .\label{theta}
\end{equation}
with $\theta_{0}$ being its value at  the initial configuration: 
\begin{equation}
    \cos(\theta_0) \coloneqq \frac{9b+27\bar{\rho}_0-2}{2\sqrt{(1-3b)^3}}\, .
\end{equation}
Likewise, the energy density of the brane evolves as
\begin{equation}
    \quad\quad  \bar{\rho}(\theta)\, =\, \frac{2}{27}\left[1+\sqrt{(1-3b)^3}\cos(\theta)\right]-\frac{b}{3}\, . \label{EnergyDensity1}
\end{equation}
From Eqs.~(\ref{lowenergy1})-(\ref{lowenergy3}) it is clear that the first and the last solutions, i.e. $\bar{H}_{1}(\theta)$ and $\bar{H}_{3}(\theta)$, are negative. However, in the range $0<\theta \leq\theta_0$,  $|\bar{H}_{3}(\theta)|$ gives a descending Hubble rate  from  its initial condition at $\theta_0$ toward $\theta=0$ (cf. dashed blue curve in the right panel of Fig.~\ref{Fig:LowEnergyRegime}); thus it cannot be a suitable solution for the collapsing process. In contrary, the solution $|\bar{H}_{1}(\theta)|$ is  ascending in the range $\theta=\theta_0$ and $\theta=0$ (cf. solid blue curve in the right panel of Fig.~\ref{Fig:LowEnergyRegime}), so it illustrates a collapse scenario. Accordingly, the second solution, i.e. $\bar{H}_{2}(\theta)$, is always positive  performing an expanding phase which is not of our interest in this range (cf. dashed red curve in the right panel of Fig.~\ref{Fig:LowEnergyRegime}).

Following the above arguments, we have a single physically reasonable solution which is given by Eq.~(\ref{lowenergy1}). This solution should respect the energy condition  (\ref{NEC-negative}), i.e., for all $\theta$ this solution should satisfy $\bar{H}_1(\theta)<\bar{H}_1(\theta_0)<\bar{\mathscr{H}}_2^{(-)}$, thus 
\begin{equation}
  \quad \ \ \cos\left(\frac{\theta_0+\pi}{3}\right)>\cos\left(\frac{\theta+\pi}{3}\right)>\cos\left(\frac{2\pi}{3}\right).
\end{equation}
This gives a condition on the brane energy density $\bar{\rho}_{-}<0<\bar{\rho}_0 <\bar{\rho}$ which is already satisfied. Consequently, $\bar{\rho}_0$ implies 
\begin{equation}
   \quad  \cos(\theta_0) > \frac{9b-2}{2\sqrt{(1-3b)^3}} =: \cos(\theta_{\rm max})\, ,
\end{equation}
thus, $0<\theta \leq\theta_0<\theta_{\rm max}$.
\item \label{limiting-item}  Once the energy density of the brane reaches the value $\bar{\rho}_{+}$, the discriminant function $N$ vanishes and the collapse gets to the limiting regime. There exist  two real solutions in this limit as
\begin{align}
\quad \quad \bar{H}_{1} &=- \frac{1}{3}\left(\sqrt{1-3b}+1\right) = \bar{\mathscr{H}}_1^{(-)},
\label{limitingone}
\\
\bar{H}_{2} &= \frac{1}{3}\left(2\sqrt{1-3b}-1\right),
\label{limitingtwo}
\end{align}
which are constants.
The solution $H_{2}$ above is  negative being the limit of Eq.~(\ref{lowenergy1}) when $\theta \rightarrow 0$.
On the other hand, $H_{2}$ is always positive being the limit of Eq.~(\ref{lowenergy2}), thus it is not associated with a collapse scenario. 
\item\label{HighEnrgy-item} If the energy density of the brane continues growing, the collapse then enters the  high energy regime (where $\bar{\rho}>\bar{\rho}_{+}$). In this regime $N>0$, so that we  have only one real solution  for the Hubble rate. This solution is given by Eq.~(\ref{EXP1}) for $\eta>0$. Subsequently, since this solution is always positive, it cannot be a reasonable candidate for the Hubble rate.  
\end{enumerate}
In summary, in the range $0< b< 1/4$, the  physically relevant solutions for the collapse on the brane is provided by Eq.~(\ref{lowenergy1}) in the low energy regime and Eq.~(\ref{limitingone}) in the limiting regime. 

\subsection{The case $\frac{1}{4}\leq b<\frac{1}{3}$}
\label{B}

In this range, since $N>0 $, there exists a unique real solution for $\bar{H}$ which is described by the  Eq.~(\ref{EXP1}). Therefore, this range of $b$ does not provide a physically suitable solution to the collapse scenario.

\subsection{The case $b=\frac{1}{3}$}
\label{C}

For this value of $b$, the densities (\ref{rhoone}) and (\ref{rhotwo}) are equal and become  $\bar{\rho}_+=\bar{\rho}_-=-1/27<0$. Therefore, $N>0$ and there is a unique solution which is given by
\begin{equation}
\bar{H}_1(t)\ = \frac{1}{3}\left[(1+27\bar{\rho})^{1/3}-1\right].
\label{b=1/3case}
\end{equation}
This solution is always positive and hence  inappropriate for describing our desirable contracting phenomenon.

\subsection{The case $b>\frac{1}{3}$}
\label{D}

In this case $\rho_+ $ and $\rho_- $ are complex conjugates which implies that $N>0$. Thus, the solution reads as
\begin{equation}
\bar{H}_1(\vartheta) =\frac{1}{3}\left[2\sqrt{3b-1}\sinh\left(\frac{\vartheta}{3}\right)-1\right],
\label{Hubble-extra1}
\end{equation}
where, $\vartheta$  is defined by
\begin{equation}
\quad  \sinh(\vartheta):= \frac{S}{\sqrt{Q^3}}~, \quad \quad \cosh(\vartheta):=
\sqrt{\frac{N}{Q^3}} \, .
\label{Hubble-extra2}
\end{equation}
Since $\vartheta$ is always positive, the above solution takes always  positive values. Therefore, we discard this solution.

For the sake of completeness in this section, let us analyze the  collapse in the standard DGP model.   For this case, no GB component is present on the (Minkowski) bulk, so by setting $\alpha=0$ in the Friedmann equation (\ref{modifiedfriedmann}), for a normal branch,  we get
\begin{equation}
H^2\, =\, \frac{\kappa_{(5)}^2}{6r_c}\rho -
\frac{H}{r_c}\, .
\label{modifiedfriedmann-DGP}
\end{equation}
This is a quadratic equation for $H$ which has only one negative solution:
\begin{equation}
H(\rho) = -\frac{1}{2r_c}\left(1+\sqrt{1+\frac{4}{3}\kappa_{(4)}^2r_c^2\rho}\right).
\label{DGP-pure-sol}
\end{equation}
This solution represents a dynamical evolution of a collapse cloud, started evolving from an initial density $\rho_0>0$, toward $a=0$. Thus, at the center, both the energy  density and the Hubble rate blow up and the collapse ends up  in a {\em shell-focusing singularity}.

In the next sections  we will elaborate more on the physical implications of the  solutions (\ref{lowenergy1}) and (\ref{limitingone}) obtained for the range $0< b < 1/4$.

\section{The fate of the singularity}
\label{Sec:singularity}

In the previous section we have found the physically reasonable solution (\ref{limitingone}) to the generalized Friedmann equation (\ref{Friedmannnb}). In this section, we will look into the fate of the collapse provided by this solution and the  types of the singularities it may encounter. In particular, to make a comparison with the well-known  Oppenheimer-Snyder dust model, we will consider  the collapse matter content as a dust fluid\footnote{This point is subtle since considering a  {\em baryonic} dust content on the brane is not a viable assumption in {\em high energy regimes}. However, in order to pursue our purpose to compare the outcome of the present collapse scenario  with that  of the general relativistic dust fluid, we should narrow down our basic assumption to a dark matter setting, which can be a physically reasonable candidate for the fluid of the same feature in high energy.} with $w=0$.

We start with analyzing  the behavior of the first (comoving) time derivative  of the Hubble rate. From the Raychaudhuri equation (\ref{Eq:Raychaudhuri-1})  for the $(-)$ branch, we get
\begin{equation}
\dot{H} =   -\frac{\kappa_{(4)}^2r_c H\rho}{2r_c H + (1+8\alpha H^2)} \,  .
\label{eq24}
\end{equation}
It turns out that some interesting features of the collapse can be
achieved once the denominator of Eq.~(\ref{eq24}) vanishes:
\begin{equation}
8\alpha H^2+2r_c H + 1=0.
\label{roots}
\end{equation}
 This quadratic equation has two real roots 
\begin{align}
\mathsf{H}_1&=  -\frac{r_{c}}{8\alpha}\left(1+ \sqrt{1-3b}  \right), \label{fate-HB1}\\
\mathsf{H}_2 &=  -\frac{r_{c}}{8\alpha}\left(1- \sqrt{1-3b}  \right). \label{DvrgDimsnless}
\end{align}
These roots are in fact equivalent respectively, to the lower and the upper bound (\ref{EC2b0}) and (\ref{EC2b}) of the $\bar{H}$ provided by the NEC (\ref{NEC-negative}).
The only physically relevant root of Eq.~(\ref{roots}) which adjusts to the   solution (\ref{lowenergy1}) is $\mathsf{H}_{1}$, i.e., the limiting solution; $\bar{\mathsf{H}}_1\equiv b r_c \mathsf{H}_{1}= \bar{\mathscr{H}}_{1}^{(-)}$. 

The solution (\ref{fate-HB1}) indicates that, for an initial configuration  in the low energy regime, as the dust cloud collapses toward the limiting regime, through the solution (\ref{lowenergy1}), the first time derivative of the Hubble rate diverges, whereas, the energy density and the Hubble rate itself remain finite: 
\begin{equation}
    \theta\to 0:\quad  \bar{\rho}\rightarrow \bar{\rho}_+,  \quad \bar{H}_1(\theta)\rightarrow \bar{\mathscr{H}}_{1}^{(-)},  \quad \dot{\bar{H}}_1\rightarrow \infty .
    \label{suddenSing}
\end{equation}
Consequently, the effective equation of state (\ref{EoS-eff}) also converges as
\begin{eqnarray}
w_{\rm eff} = -1 + \frac{2r_cH}{2r_cH + (1+8\alpha	H^2)}\frac{\rho}{\rho_{\rm eff}}  \, ,
\label{EoS-eff2}
\end{eqnarray}
where $\rho$ and $\rho_{\rm eff}$ remain finite.
Therefore, a peculiar abrupt event occurs in a finite (comoving) time $t_{\rm sing}$ in the limiting regime, prior to the formation of the  shell-focusing singularity at $R=ra=0$. 
In a (late-time) cosmological setting, such an abrupt event  is  called a {\em sudden  singularity} \cite{Barrow:2004xh,FernandezJambrina:2004yy, BouhmadiLopez:2009jk}. Following such denomination, we thus call it, within the present setting, a {\em sudden black hole/naked singularity} depending on  whether or not the singularity will be trapped by  an apparent horizon at the final stages of the  collapse.

\begin{figure}
	\begin{center}
		\includegraphics[scale=0.45]{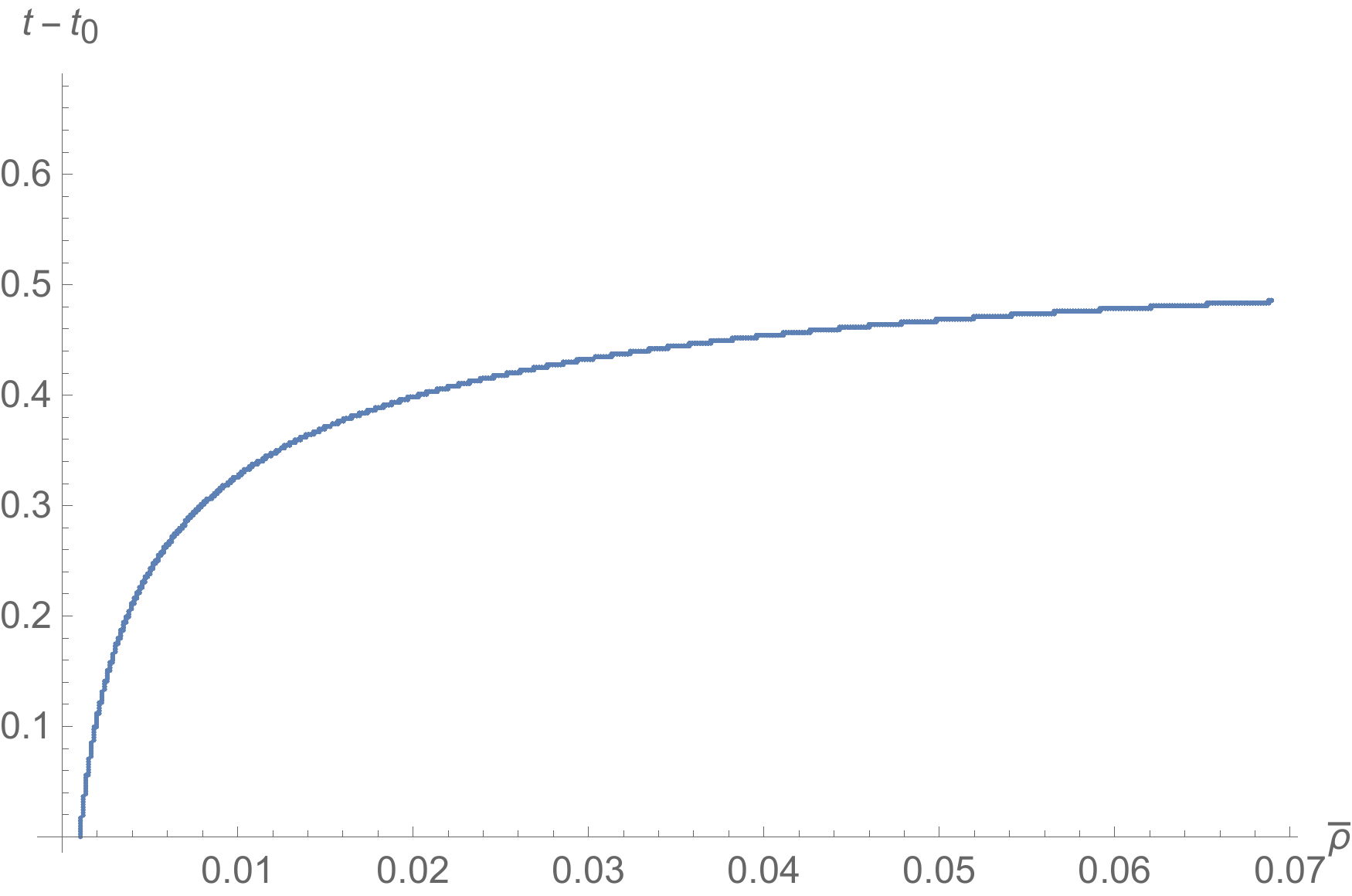} 
		\caption{Variation of the comoving time $t-t_0$ in terms of the energy density $\bar{\rho}$ of the brane. The time left before the brane hits the sudden singularity, $t_{\rm sing}-t_0$, is shown at the singular energy density $\bar{\rho}_+$. For fixed values of $r_c=0.2$ and $b=1/8$, we get $\bar{\rho}_+\approx 0.069$ and $t_{\rm sing}-t_0\approx 0.45$.} \label{Fig-time}
	\end{center}
\end{figure}

Finally, using the relation between the Hubble parameter
and the energy density, $\dot{\rho}+3H\rho=0$, one can write the Hubble rate as a function of  comoving time. Then, by integrating this equation, the time remaining before the dust brane hits the sudden singularity is obtained as
\begin{equation}
   t_{\rm sing} - t_0 = -\frac{br_c}{3}\int_{\bar{\rho}_0}^{\bar{\rho}_{+}} \frac{d\bar{\rho}}{\bar{\rho}\, \bar{H}(\bar{\rho})}\, ,
   \label{time-sing}
\end{equation}
where, $t_0$ and $t_{\rm sing}$ denote the
present time and the time at the sudden singularity, respectively.
To evaluate the integral (\ref{time-sing}), it is convenient to rewrite the solution (\ref{lowenergy1})  explicitly as a function of $\bar{\rho}$ as
\begin{align}
\bar{H}_1(\bar{\rho}) &= -\frac{1}{2}\left(s_{+} + s_{-}\right)   - \frac{i\sqrt{3}}{2}\left(s_{+}-s_{-}\right) - \frac{1}{3}\, , \quad \label{h-eff2}
\end{align}
where, 
\begin{eqnarray}
s_{\pm}(\bar{\rho})\ :=\  \left[S(\bar{\rho})\pm\sqrt{Q^3+S^2(\bar{\rho})}\right]^{\frac{1}{3}} .
\end{eqnarray} 
The result is shown in Fig.~\ref{Fig-time}, where $t_{\rm sing}-t_{0}$ is evaluated for different values of the energy density in the range $\bar{\rho}_0<\bar{\rho}<\bar{\rho}_{+}$.

As a specific case, given by the solution (\ref{DGP-pure-sol}) for the standard DGP model, the time derivative of the Hubble rate is derived from the Friedmann equation (\ref{modifiedfriedmann-DGP}) as
\begin{equation}
    \dot{H} = \frac{\kappa_{(4)}^2 H \rho}{2H+1/r_c}\, .
\end{equation}
This is equivalent to the Eq.~(\ref{eq24}) for $\alpha=0$ which implies that, for the case $H=-1/2r_c$, the time derivative of Hubble parameter diverges. This value for  $H$  does not display  a physically relevant situation, i.e.,  matching with the solution (\ref{DGP-pure-sol}). Therefore, only a shell-focusing singularity at $R=0$ can occur in this case.

\section{Trapped surfaces and the exterior geometry}
\label{Sec:Exterior-geometry} 

To complete our model of gravitational collapse in DGP-GB brane scenario, we need to match the interior spacetime (\ref{metric-interior}) to a suitable exterior geometry\footnote{Because of the extradimensional effects, i.e., the confinement of the dust cloud  to the brane while the gravitational field can access the extra dimension, and  the non-local gravitational interaction between
the brane and the bulk, the standard 4D matching conditions (cf. \cite{Israel:1966a,Israel:1966b}) on the brane  are much more complicated to implement \cite{Germani:2001du,Visser:2002vg}. However,  we have  ended up with a modified matter collapse (different from the standard dust matter), with an effective energy density (\ref{energydensity-1}) and  pressure (\ref{pressure-eff}) satisfying the standard Friedman equation. Thus, from an effective point of view (on the brane), we can  employ the 4D matching at the  2-surface boundary of this peculiar matter cloud and look for a suitable exterior geometry (see \cite{Bruni:2001fd, Tavakoli:2015llr} for a similar approach).}. 
In an advanced Eddington-Finkelstein-like coordinates $(v, r_{v})$, a general  exterior metric is  
written as \cite{Joshi:2008zz}:
\begin{equation}
d\tilde{s}^2=-f({v},r_{v})\, d{v}^2-2dv\, dr_{v}+r_{v}^2\, d\Omega^2.
\label{metric2}
\end{equation}
In this coordinates system, the {\em trapping horizon} is given simply by the relation $f({v},r_{v})=0$.
In general,  the formation of the (shell-focusing) singularity is independent of matching the interior and the exterior spacetimes.
For GR settings, when the interior matter content is a pressureless dust fluid, an exterior Schwarzschild geometry can be matched at the boundary of the two regions \cite{Hawking:1973uf, Oppenheimer:1939ue}. However, in the herein effective theory, due to  presence of an effective nonzero pressure (\ref{pressure-eff}), induced by the DGP-GB geometrical terms, the exterior region cannot be a Schwarzschild spacetime. It follows that, a more realistic picture  should involve a radiation zone and matter emissions in the outer region. Therefore, we match the interior, at the boundary $\Sigma$ of the dust cloud, to an exterior  geometry described by a general class of non-stationary metric, called the {\em generalized Vaidya spacetime} \cite{Vaidya:1999zz}. Its geometry is described by the metric (\ref{metric2}) where the boundary function $f(v, r_v)$ is defined as
 \begin{equation}
f(v, r_v) := 1-\frac{2GM({v},r_{v})}{r_{v}}\, .
\label{Vaidya}
\end{equation}
Here, $G$ is the $4D$ Newton's constant given by the term  $\kappa_{(4)}^2=8\pi G$, and $M$ is called the {\em generalized Vaidya mass} which depends on  the retarded null coordinates $(v,r_{v})$.

According to GR, the matching conditions allow 
us to study the  formation of apparent horizons that we refer to it later. Thereby, in the herein effective model for the collapse of a fluid, having the effective density and pressure $\rho_{\rm eff}$ and $p_{\rm eff}$,  the Israel-Darmois  junction conditions \cite{Israel:1966a,Israel:1966b} should be satisfied at the boundary $\Sigma$  of the fluid (with the boundary shell $r = r_{b}$). 
This implies that, at $\Sigma$, two fundamental forms, namely, the metric and the extrinsic curvature  must match. 
Matching the area radius at the boundary results in $R(t, r_b)=r_ba(t)=r_v$. Likewise,  matching  the first and second fundamental forms on $\Sigma$, for the FLRW interior metric and the generalized Vaidya exterior metric, leads to \cite{joshi_2007}:
\begin{align}
F_{ \rm eff}(t, r_{b})\ &=\ 2GM({v},r_{v}),
\label{V2}
\\
\left(dv/dt\right)_{\Sigma}\ &=\ (1+r_{b}\dot{a})\left(1-\frac{F_{\rm eff}}{R}\right)^{-1}\ ,
\label{V3-a}
\\
GM({v},r_{v})_{,r_{v}}\ &=\ \frac{F_{\rm eff}}{2R}+r_{b}^2a\ddot{a},
\label{V4-a}
\end{align} 
where, $F_{\rm eff}(t, r_b)$ is interpreted as the {\em effective} Misner-Sharp mass function \cite{misner1964relativistic}, being the total effective mass within the boundary shell labelled by $r_b$ at time $t$, which satisfies the relation
\begin{eqnarray}
F_{\rm eff} \ =\ \frac{\kappa^2_{(4)}}{3}\rho_{\rm eff}R^3 .
\label{EffectiveMass}
\end{eqnarray}
Note that, in the GR context,  $\rho_{\rm eff}$ is replaced by $\rho$, so that the effective mass function becomes the usual gravitational mass $F=(\kappa^2_{(4)}/3)\rho R^3$. For the dust cloud within $r_b$, by setting $\rho=\rho_0(R_0/R)^3$ (where $R_0\equiv r_ba_0$ is the physical radius of the boundary shell), this yields 
\begin{equation}
    F=2Gm_0 = const.\, ,
    \label{mass-star}
\end{equation}
where $m_0$ is the total mass of the star (i.e., the physical mass contained inside  $R_{0}$).

\subsection{Evolution of trapped surfaces}

\begin{figure*}[!ht]
	\begin{center}
		\includegraphics[scale=0.4]{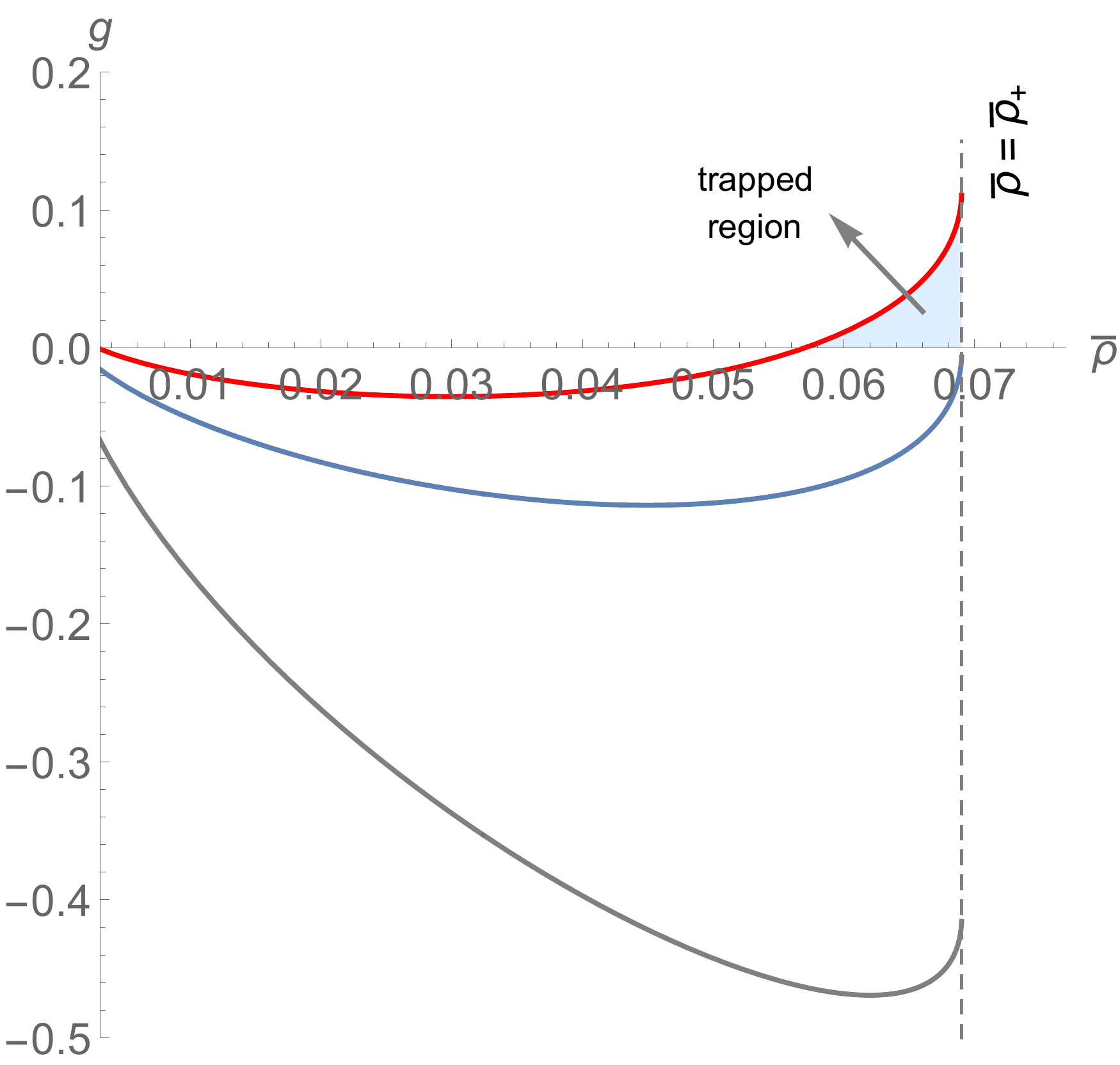} \quad \quad  \quad 
			\includegraphics[scale=0.48]{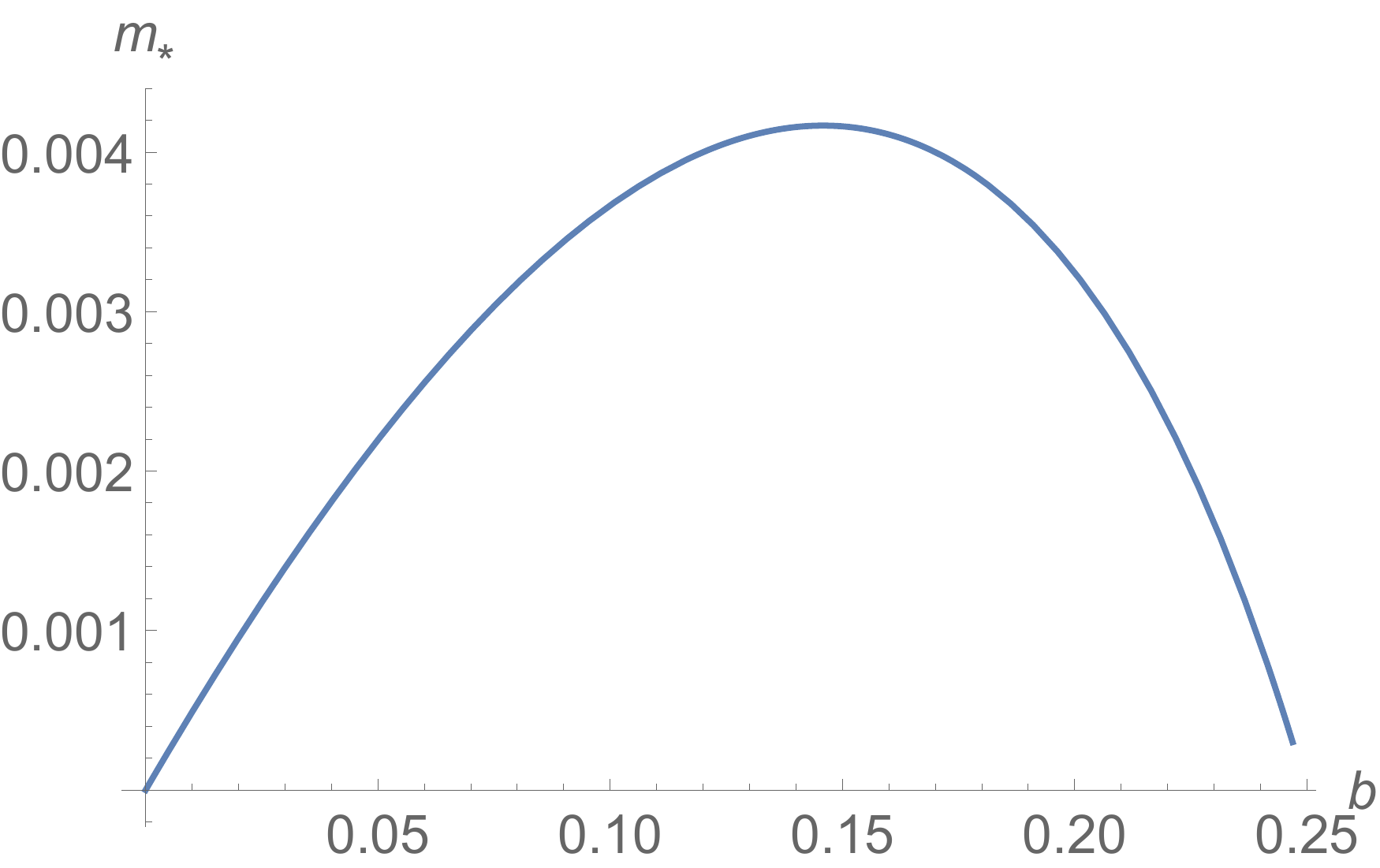} 
		\caption{Left: the evolution of $g(\bar{\rho})$ is plotted for different initial conditions $R_0,~\bar{\rho}_0$ and for the given model parameters $b=1/8,~ r_c=0.2$. It is shown that, for the specific initial conditions, no  event horizon  forms during the collapse prior to the occurrence of the sudden singularity at $\bar{\rho} = \bar{\rho}_{+}\approx 0.069$. Right: behavior of the threshold mass $m_\ast$ is plotted with respect to the parameter $b$ for the fixed value of $r_c=0.2$.} \label{Fig-mass-star}
	\end{center}
\end{figure*}

The  geometry of the trapped surfaces is a key to ascertain the formation of a black hole or a naked singularity. The ratio of the  effective mass to the physical radius of  the collapse $R$ determines the properties  of the trapped surface. When the  mass function satisfies the inequality $F_{\rm eff}<R$ no trapped surfaces would form during the dynamical evolution of the collapse, whereas for the case $F_{\rm eff}\geq R$  trapped surfaces will emerge.
We now proceed by studying the behavior of 
$F_{\rm eff}$, evaluated at the  solution (\ref{lowenergy1}). 
By substituting the effective energy density (\ref{energydensity-1}), for the $(-)$ branch, into Eq.~(\ref{EffectiveMass})  we obtain
\begin{align}
F_{\rm eff} &= F -
\frac{1}{r_c} \left(1+\frac83\alpha H^2\right)HR^3 = 2GM.
\label{massF-eff}
\end{align}
To examine the visibility of the sudden singularity, we rewrite the condition $F_{\rm eff}<R$  in a dimensionless form  
\begin{align}
g(\bar{\rho}) := \bar{\rho}-
\left(b+\bar{H}^2\right)\bar{H}  - \bar{H}_h^2<0\, ,
\label{horizon-condition0}
\end{align}
where
\begin{align}
\bar{H}_h(\bar{\rho}) :=  -\frac{br_c}{R_0} \left(\frac{\bar{\rho}}{\bar{\rho}_0}\right)^\frac{1}{3}.
\label{horizon-condition00}
\end{align}
The Hubble rate $\bar{H}$ in Eq.~(\ref{horizon-condition0}) is the solution to the Friedmann equation (\ref{Friedmannnb}). So, the terms $\bar\rho - (b + {\bar H}^2)\bar H$  can be replaced by ${\bar H}^2$. This  yields 
\begin{eqnarray}
g(\bar{\rho}) := \bar{H}^2(\bar{\rho}) - \bar{H}_h^2(\bar{\rho})<0,  \quad \text{for $\bar{\rho}_0\leq\bar{\rho}\leq\bar{\rho}_{+}$}. \quad \quad 
\label{horizon-condition}
\end{eqnarray}
This inequality implies that, to avoid the trapped surfaces forming,  for all $\bar{\rho}_0\leq\bar{\rho}\leq\bar{\rho}_{+}$, the value $|\bar{H}_1(\bar{\rho})|$  should be {\em always} smaller than $|\bar{H}_{h}(\bar{\rho})|$, i.e. $\bar{H}_{h}(\bar{\rho})<\bar{H}_1(\bar{\rho})<0$, from the initial configuration up to the formation of the sudden singularity.
In the case where (\ref{horizon-condition}) does not hold, i.e., $g(\bar{\rho})\geq 0$, trapped surfaces would form and the sudden singularity  will be covered by the black hole horizon. The equality gives the energy density $\bar{\rho}_h$ or the physical radius $R_h$, at the apparent horizon: 
\begin{equation}
    \bar{H}_1(\bar{\rho}_h)  =  -\frac{br_c}{R_0} \left(\frac{\bar{\rho}_h}{\bar{\rho}_0}\right)^\frac{1}{3} \quad  \text{or} \quad  \bar{H}_1(R_h)=  -\frac{br_c}{R_h}\, .
\end{equation}
Once the energy density $\bar{\rho}_h$ is found from the left equation above, the time during which the horizon forms, i.e. $t_h-t_0$, can be determined from the figure~\ref{Fig-time}.

Figure~\ref{Fig-mass-star} depicts the condition (\ref{horizon-condition})   for the evolution of the apparent horizon obtained for different values of initial conditions $R_0$ and $\bar{\rho}$. The ``red curve'' represents an initial condition for which an apparent horizon forms before the brane reaches to the sudden singularity. The ``gray curve'' shows a condition for the formation of a {\em naked} sudden singularity, and the ``blue curve'' represents a limiting situation, where a horizon forms at the singular epoch. 
It turns out that, if (\ref{horizon-condition}) holds initially, i.e., $g(\bar{\rho}_0)<0$, then to ensure that trapped surfaces will never evolve during the collapse,  $g(\bar{\rho})$ would vanish only where $\bar{\rho}>\bar{\rho}_{+}$. Let us be more precise as follows. Suppose that   (\ref{horizon-condition}) holds at the initial configuration. Accordingly, this gives the condition 
\begin{align}
 	 H_0>-\frac{1}{R_0}\, ,
 	\label{EC2b-2}
\end{align}
between the initial Hubble rate $\bar{H}_0\equiv\bar{H}_1(\bar{\rho}_{0})$ and the initial physical radius $R_0$ of the dust.
Then, the necessary and sufficient condition for formation of trapped surfaces during the collapse is that  $g(\bar{\rho}_{+})\geq0$. This yields 
\begin{equation}
    \frac{br_c}{R_0} \left(\frac{\bar{\rho}_{+}}{\bar{\rho}_0}\right)^\frac{1}{3}\leq\frac{1}{3}\left(\sqrt{1-3b}+1\right). \label{horizoncondition2}
\end{equation}
Now, by substituting  $\bar{\rho}_{+}$ from Eq.~(\ref{rhoone}) into the relation above,  after some simplification,  Eq.~(\ref{horizoncondition2}) becomes
\begin{equation}
  m_0\geq m_\ast\, , \label{thresholdmass}
\end{equation}
where,  $m_{\ast}$ is a constant, with the dimension of mass, defined by
\begin{equation}
  m_\ast\, :=\, 
    \frac{br_c}{2G}\frac{2\left[1+\sqrt{(1-3b)^3}\right]-9b}{\left(\sqrt{1-3b}+1\right)^3}\, ,
\end{equation}
which explicitly depends  only on the DGP-GB  parameters $b$ and $r_c$. The right plot in the Fig.~\ref{Fig-mass-star} represents the behavior of the $m_\ast$ with respect to $b$ in the range $0<b<1/4$ for a fixed value of $r_c$. 
Clearly, the parameter $m_\ast$ represents a threshold mass for the black hole formation. In other words, if the initial mass $m_0$ of the  dust cloud  is smaller that $m_\ast$, then no horizon would form during the collapse up to the sudden singularity. But, if $m_0\geq m_\ast$, then formation of trapped surfaces will be ensured and  the final sudden singularity will be hidden behind a black hole horizon.

For the standard DGP model, with the solution (\ref{DGP-pure-sol}), the effective mass function (\ref{massF-eff}) reduces to
\begin{align}
F_{\rm eff} = F +\frac{1}{2r_c^2}\left(1+\sqrt{1+\frac{4}{3}\kappa_{(4)}^2r_c^2\rho}\right)R^3\, . 
\label{massF-eff-DGP}
\end{align}
This expression is always positive so that $F_{\rm eff}>R$, which implies that the singularity will be hidden behind the event horizon of a black hole.
To be more precise, let us rewrite the condition (\ref{horizon-condition}) for the solution (\ref{DGP-pure-sol})  as
\begin{eqnarray}
m_0 < - \frac{4\pi\rho}{3H^3(\rho)}\,  , \quad  \quad \text{for all~ $\rho\geq\rho_0$}. \quad 
\label{horizon-condition-DGP}
\end{eqnarray}
If this condition holds initially, in order to avoid the horizon formation, the dust cloud should remain untrapped until when the singularity at $R=0$ is reached. It follows that, in the singular limit, i.e.,  as $R\rightarrow0$, where the dust density blows up as $\rho\rightarrow\infty$, the right-hand side of the inequality (\ref{horizon-condition-DGP}) vanishes:
\begin{align}
 -\frac{4\pi\rho/3}{H^3(\rho)}\, \approx\,  \frac{1}{\rho^{1/2}} \rightarrow 0 \qquad \Rightarrow \qquad m_0<0. \quad \quad 
\label{horizon-condition-DGP2}
\end{align}
This implies that, for  condition (\ref{horizon-condition-DGP}) to be satisfied during the collapse, the star mass $m_0$ should be negative in the vicinity of the singularity $R=0$, which cannot be true. Therefore, for all initial star mass $m_0>0$, in the standard DGP model, formation of the black hole   is inevitable.

\subsection{The exterior geometry}
	
\begin{figure}
	\begin{center}
	\includegraphics[scale=0.45]{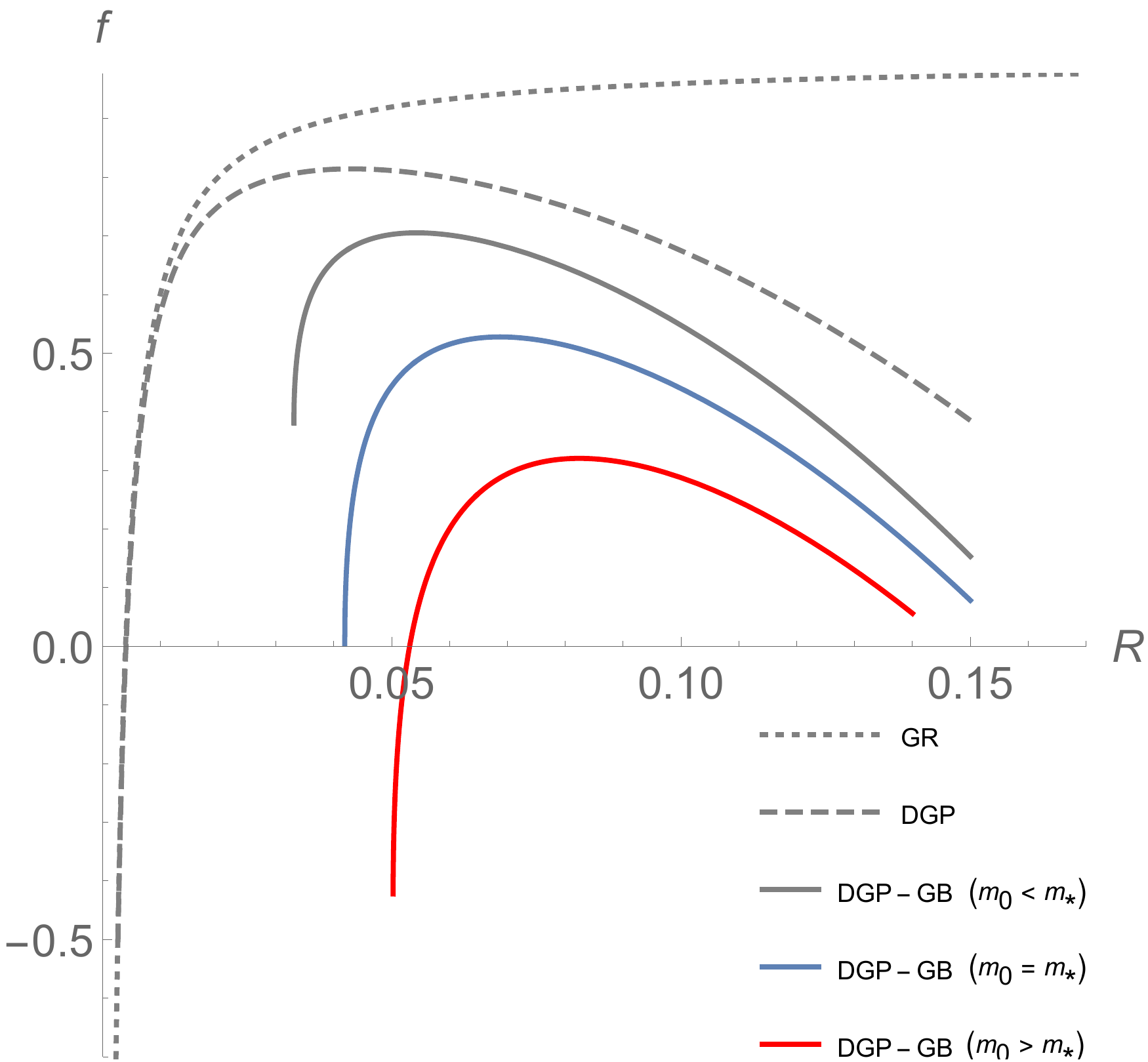}
		\caption{Behavior of the exterior function $f(R)$ for different values of the star mass $m_0$ with respect to the threshold mass $m_\ast\approx 0.004$. The physical parameters are fixed for $b=1/8$, $r_c=0.2$ and $G=1$.
		The dotted   curve matches the  Schwarzschild geometry (i.e. the GR limit). The dashed curve depicts a black hole geometry in the standard DGP model [cf. Eq.~(\ref{exterior-DGP})]. The solid curves represent the results in the DGP-GB model (i.e. Eq.~(\ref{sol-ext1})) for different ranges of the star mass $m_0$: the gray curve accords with a spacetime with no horizon. The red curve is associated to a black hole spacetime where the sudden singularity is trapped inside the horizon. The limiting condition is depicted by the blue curve where the horizon forms at the sudden singularity.} \label{Fig-Exterior}
	\end{center}
\end{figure}

The matching conditions (\ref{V2})-(\ref{V4-a}) lead to the exterior geometry (\ref{metric2}) with  the boundary function
\begin{align}
f(R) &= 1-\frac{F_{\rm eff}}{R} = 1-\frac{\kappa_{(4)}^2}{3}\rho_{\rm eff}R^2 \nonumber \\
&   = 1-H^2R^2 .
\label{exterior}
\end{align}
It is clear that, on the onset of  the collapse, the  condition (\ref{EC2b-2}) for the absence of trapped surfaces (in the interior region) can be translated  to the condition $f(R_0)>0$ on the boundary function (in the exterior geometry).
Now, for the interior solution $\bar{H}_1(\bar{\rho})$, we get
\begin{align}
f(R)  =  1-\frac{2GM(R)}{R}   = 1-\left(\frac{\bar{H}_{1}R}{br_c}\right)^2 ,
\label{exterior0}
\end{align}
where $M(R)$ is the effective mass of the exterior geometry.
The  physical solution $\bar{H}_1$  should be written now in terms of the exterior parameters (i.e., $m_0$ and $R$). 
Given the solution (\ref{h-eff2}), the expression for $S(\bar{\rho})$ in terms of the exterior parameters can be rewritten 
from Eq.~(\ref{eq11})   as
\begin{eqnarray}
S(R)\ =\ \frac{b}{6}-\frac{1}{27}+b^2r_c^2\frac{Gm_0}{R^3}\, .
\end{eqnarray}
Now, in terms of the parameters $m_0, b$, and $r_c$, we can 
write  the  solution $f(R)$ as
\begin{align}
f(R) &= 1 - \frac{R^2}{b^2r_c^2}\Bigg\{ \frac{1+i\sqrt{3}}{2}\Bigg[\left(\frac{b}{6}-\frac{1}{27} + \frac{b^2r_c^2Gm_0}{R^3}\right)\nonumber \\ 
&  \quad +\sqrt{Q^3+\left(\frac{b}{6}-\frac{1}{27} + \frac{b^2r_c^2Gm_0}{R^3}\right)^2}\Bigg]^{\frac{1}{3}} \nonumber \\ 
& \quad  
+\frac{1-i\sqrt{3}}{2}\Bigg[\left(\frac{b}{6}-\frac{1}{27} + \frac{b^2r_c^2Gm_0}{R^3}\right)\nonumber \\
& \quad -\sqrt{Q^3+\left(\frac{b}{6}-\frac{1}{27} + \frac{b^2r_c^2Gm_0}{R^3}\right)^2}\Bigg]^{\frac{1}{3}}  + \frac{1}{3}\Bigg\}^2 .  
\label{sol-ext1}
\end{align}
The behavior of $f(R)$ is plotted in  Fig.~\ref{Fig-Exterior}. It is shown that, there exist three different cases depending on the initial mass $m_0$: If $m_0<m_\ast$ (gray solid curve), then $f(R)$ is always positive which represents a spacetime geometry with no horizon. On the other hand, for $m_0>m_\ast$ (red  curve), there exists a trapped region, where $f(R)<0$, which covers the sudden singularity. This represents a black hole geometry. Finally, the case $m_0=m_\ast$ represents a limiting condition for the formation of the black hole. In other words, in this case the black hole horizon forms only on the sudden singularity.

Finally, by substituting  the original  DGP solution (\ref{DGP-pure-sol}) into (\ref{exterior}), the boundary function  becomes
\begin{align}
f(R) &=  1-\frac{2G m_0}{R} \nonumber \\ & \quad\quad  -\frac{R^2}{2r_c^2}\left(1+\sqrt{1+\frac{8r_c^2 Gm_0}{R^3} }\right). 
\label{exterior-DGP}
\end{align}
This function represents  a black hole spacetime whose geometry is different from the Schwarzschild one, the difference being involved in the last term (cf. the dashed curve in Fig.~\ref{Fig-Exterior}). As it is clear  from the Fig.~\ref{Fig-Exterior}, at high energies in the vicinity of the general relativistic singularity (i.e., in the limit $R\ll r_c$), the second term in Eq.~(\ref{exterior-DGP}) is negligible so that the standard DGP model becomes purely 4D, while  at low energies (i.e., $R\gg r_c$) the second term  becomes dominant. This is a consequence of the fact that, the original  DGP model provides only  IR modifications to the brane and cannot change the nature of the UV regime.

In addition, in Fig.~\ref{Fig-Exterior}, the boundary function $f(R)$ in the GR and the standard DGP settings were plotted for the mass $m_0=0.002<m_\ast$. Clearly, for this range of star mass, trapped surfaces  form as the collapse proceeds. However, the GB modification to the high energy regime, prevents the formation of the horizons by setting a threshold mass $m_\ast$ for the horizon to develop.

\section{Conclusions and Outlook}
\label{conclusion}

In this paper, we have 
investigated the gravitational collapse of a dust cloud in an induced gravity model, specifically where a dust fluid is present on a DGP brane, whereas a  GB term is provided for the bulk. In particular, by considering a normal (non self-accelerating) DGP branch, we studied the evolution equation for the collapse and derived different classes of solutions.

Concretely,  by applying suitable  physical conditions on these solutions (i.e., qualifying a contracting branch governing the collapse dynamics, and  the required energy conditions),
we obtained a {\it new} collapse scenario with a {\em distinct} set of characteristic features. To be more precise, this scenario represents a process in which the brane starts its evolution from an appropriate initial condition toward  vanishing physical radius $R=0$, where the general relativistic (shell-focusing) singularity is located. However, prior to the formation of this shell-focusing singularity, the brane reaches a regime where the Hubble rate and the energy density get  finite specific values, while the (comoving) time derivative of the Hubble rate diverges.
Aware of 
a labeling chosen for a phenomenon in late-time cosmological setting, we analogously suggest herein to call this abrupt event a {\em sudden naked/black hole singularity}, depending on the avoidance/formation of the apparent horizon.

Subsequently, we studied the evolution of the trapped surfaces on the brane. 
In doing so, we aimed at establishing whether the specific IR-UV combination employed within this manuscript, would allow new features regarding the  CCC. In particular, whether a GB sector would  introduce any interesting feature.
We  found that, the formation of trapped surfaces depends only on the initial value of  the star mass $m_0$. Specifically,  there exists a {\em threshold mass} $m_\ast$ below which no horizon can form. 
This is indeed a consequence of the presence of the GB  term in the action, implying significant modifications at high energies. 
In particular, from further utilizing   appropriate junction conditions  on the 2-boundary  surface of the dust cloud, we obtained a suitable exterior geometry which matches to the interior region at the boundary surface. It turns out  that this exterior geometry represents a black hole only if $m_0$ is larger or equal to  $m_\ast$. Otherwise, the {\em sudden singularity} will be visible to the distant observers through the exterior region. 
It is worthy of note that, in the absence of the GB term on the bulk (i.e., for the standard DGP model), as the collapse proceeds, the formation of the  trapped surfaces  is inevitable. 
Meanwhile, as the brane  enters the  high energy regime--where  gravity becomes 4D on the brane--the induced gravity alone  does not change the process of the  singularity formation at $R=0$; thus, the final shell-focusing singularity will be covered by a black hole horizon.

Gravitational waves can provide an observational source to test some of the novel predictions and modifications to GR at high energies.
The herein DGP-GB braneworld scenario introduces significant changes to the dynamics of the gravitational
collapse and perturbations thereon, offering interesting and potentially testable implications for high energy astrophysics. 
In particular, from the brane viewpoint, the bulk effects, i.e. the high-energy  corrections and the  
corresponding Kaluza-Klein modes, act as source terms for the brane perturbations on the interior spacetime. These effects can  be carried out to the exterior spacetime geometry of the emergent (non-Schwarzschild) black hole/naked singularity through the matching conditions on the 2-boundary surface. Consequently, probing such effects through observations of gravitational waves could eventually  reveal nonstandard features of the spacetime in the high energy regime.

Alongside the collapse process due to huge distributions of matter on the brane we presented herein this paper, there is another, much more speculative, process by which a black hole (or a naked singularity) may have been produced from gravitational collapse of regions of enhanced density within the high energy regimes in the early universe. It follows that, from the GR point of view, if  sufficiently large inhomogeneities in the density were present  in the early universe, the regions of such enhanced  density could collapse to form a so-called {\em primordial black hole} rather than expand with the rest of the universe.
It turns out that, extending our herein approach by considering the whole universe on the brane (at early times), modified by a GB term,  one could explore the required conditions for formation of the primordial black holes or naked singularities. These plausible alternative features follow from assuming a suitable choice of the induced gravity and bulk parameters that can lead to an effective equation of state of dark matter.
Such study would enlarge our understanding of the issues related the dark matter and their connections to the primordial black holes/naked singularities.
This may provide an alternative answer to some of the fundamental problems in the early universe which we would like to leave for now to be addressed in our future projects.

\section*{Acknowledgments}

P.V.M. and Y.T. acknowledge the FCT grants No.~UIDBMAT/00212/2020 and No.~UIDPMAT/00212/2020 at CMA-UBI.  This article is based upon work conducted within  the  Action CA18108--Quantum gravity phenomenology in the multi-messenger approach--supported by the COST (European Cooperation in Science and Technology).

\bibliography{Bibliography}

\end{document}